\documentclass{aa}  

\usepackage{natbib}
\usepackage{graphicx}
\usepackage{txfonts}
\begin{document}

\title{Detection of 183 GHz H$_{2}$O megamaser emission towards NGC 4945}

   \author{E. M. L. Humphreys
          \inst{1}
          \and
          W. H. T. Vlemmings\inst{2}
          \and
          C. M. V. Impellizzeri\inst{3,4}
          \and
          M. Galametz\inst{1}
          \and
          M. Olberg\inst{2}
          \and
          J. E. Conway\inst{2}
         \and
          V. Belitsky\inst{2}
          \and
          C. De Breuck\inst{1}
          }

   \institute{European Southern Observatory (ESO), 
              Karl-Schwarzschild-Str. 2, 85748 Garching bei M\"{u}nchen, Germany\\
              \email{ehumphre@eso.org}
         \and
             Department of Earth and Space Sciences, Chalmers University of Technology, Onsala Space Observatory, 439 92 Onsala, Sweden
          \and
          Joint Alma Office, Alsonso de Cordova 3107, Vitacura, Santiago, Chile
          \and
          National Radio Astronomy Observatory, 520 Edgemont Road, Charlottesville, VA 22903, USA
               }

   \date{Received ; accepted }

 
  \abstract
   {}
   {The aim of this work is to search Seyfert 2 galaxy NGC 4945, a well-known 22 GHz water megamaser galaxy, for H$_{2}$O (mega)maser emission at 183 GHz.}
   {We used APEX SEPIA Band 5 (an ALMA Band 5 receiver on the APEX telescope) to perform the observations.}
   {We detected 183 GHz H$_{2}$O maser emission towards NGC 4945 with a peak flux density of $\sim$3 Jy near the galactic systemic velocity. The emission spans a velocity range of several hundred kms$^{-1}$. We estimate an isotropic luminosity of $>$ 1000 L$_{\odot}$, classifying the emission as  a megamaser. A comparison of the 183 GHz spectrum with that observed at 22 GHz suggests that 183 GHz emission also arises from the active galactic nucleus (AGN) central engine.    If the 183 GHz emission originates from the circumnuclear disk, then we estimate that a redshifted feature at 1084 kms$^{-1}$ in the spectrum should arise from a distance of 0.022 pc from the supermassive black hole (1.6 $\times$ 10$^{5}$ Schwarzschild radii), i.e. closer than the water maser emission previously detected at 22 GHz.
This is only the second time 183 GHz maser emission has been detected towards an AGN central engine (the other galaxy being NGC 3079). It is also the strongest extragalactic millimetre/submillimetre water maser detected to date.
}
   {Strong millimetre 183 GHz H$_{2}$O maser emission has now been shown to occur in an external galaxy. 
   For NGC 4945, we believe that the maser emission arises, or is dominated by, emission from the AGN central engine. 
   Emission at higher velocity, i.e. for a Keplerian disk closer to the black hole, has been detected at 183 GHz compared with that for the 22 GHz megamaser.  
   This indicates that millimetre/submillimetre H$_{2}$O masers can indeed be useful probes for tracing out more of AGN central engine structures 
   and dynamics than previously probed. Future observations
   using ALMA Band 5 should unequivocally determine the origin of the emission in this and other galaxies.}


   \maketitle
%

\section{Introduction}

Water megamaser galaxies have become the object of extensive study at 22 GHz,
since the discovery that the emission traces a sub-parsec
scale portion of the circumnuclear disk in NGC 4258, within 1 pc of
the supermassive black hole (SMBH) \citep{Miyoshi1995}.
Very Long Baseline Interferometry (VLBI) observations of the masers have provided detailed 
information on the kinematics and structure of active galactic nucleus (AGN) circumnuclear disks 
\citep[e.g.][]{Moran1995,Greenhill1997,Trotter1998,Kondratko2005,Argon2007,Humphreys2008,Braatz2010,Impellizzeri2012,Reid2013,Kuo2015,Gao2016}.
Geometric modelling of VLBI disk maser data, provided that acceleration or proper motion measurements are
also possible, can be used to perform maser cosmology and has yielded high-accuracy Hubble constant 
estimates \citep[e.g.][]{Humphreys2013,Braatz2015}. Additionally, water megamasers can originate from the interaction of 
AGN radio jets with the interstellar medium, yielding masers in shocked gas within radii 
of 1 -- 10 pc of the central regions \citep[e.g.][]{Claussen1998}.

Radiative transfer models for galactic water masers have long predicted, 
and observations have found that, the 22 GHz maser does not occur in isolation; very similar conditions to those required to produce this maser yields additional
H$_{2}$O maser lines in the millimetre/submillimetre (mm/submm) 
\citep[e.g.][]{Deguchi1977,Neufeld1991,Yates1997,Humphreys2001,Gray2016}.
It is therefore believed that some mm/submm H$_{2}$O masers could occur from broadly the same regions
as 22 GHz masers in AGN and/or could probe unchartered regions of an AGN central engine,
including regions closer to a black hole. 
Detection and study of mm/submm water masers
could therefore make unique contributions to the study of disk and radio jet structure in AGN.

\begin{figure}
   \centering
   \includegraphics[scale=0.5]{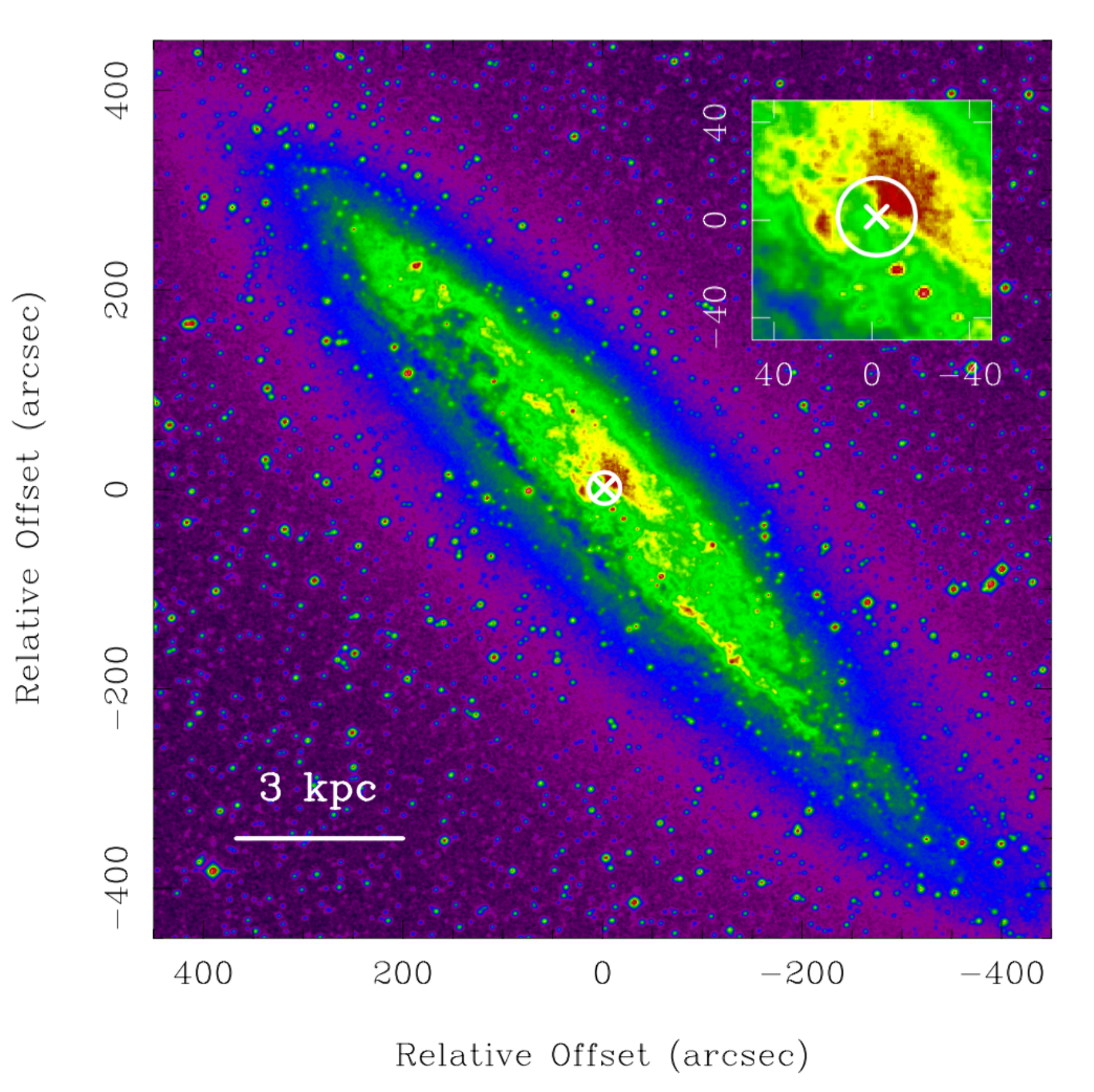}
   \caption{Image of NGC 4945 from the Palomar Sky Survey\protect\footnotemark. The cross indicates the position for NGC 4945 used in our APEX SEPIA Band 5 observations:  $\alpha_{2000}$
   =13:05:27.28, $\delta_{2000}$= $-$49:28:04.4. The circle indicates the half-power beamwidth (HPBW) of the APEX observations. At a sky
   frequency of 182.981 GHz this corresponds to 31.8\arcsec, or 570 pc at a distance of 3.7 Mpc. }
              \label{map}%
    \end{figure}

To date extragalactic mm/submm water maser emission has been detected for the 183 GHz
transition towards
NGC 3079 using the SMA \citep{Humphreys2005}, Arp 220 using the
IRAM 30 m and APEX (Cernicharo2006; Galametz et al. 2016), 
and for the 321 GHz transition towards Circinus and NGC 4945 using ALMA (Hagiwara et al. 2013; Pesce et al. 2016; Hagiwara et al. 2016). 
There has also been
 a tentative, 5$\sigma$ detection of 439 GHz H$_{2}$O water maser emission towards NGC 3079 using the James Clark Maxwell Telescope \citep{Humphreys2005}.

NGC 3079, Circinus, and NGC 4945 
display 22 GHz megamaser emission, and the mm/submm water maser emission towards
these targets has been interpreted as also arising from the AGN central engine.
However, there is no 22 GHz detection towards Arp 220 and in this case the 183 GHz water
megamaser emission is interpreted as arising from $\sim$10$^{6}$ star-forming cores \citep{Cernicharo2006}
i.e. an environment similar to that giving rise to OH megamaser emission. 

In this Letter, we report the first detection of 183 GHz H$_{2}$O megamaser emission towards southern
Seyfert 2 galaxy NGC 4945. The velocities quoted throughout this work are radio definition, local standard
of rest. The distance used is 3.7 Mpc \citep{Tully2013} and the systemic velocity adopted is 556 kms$^{-1}$, 
based on CO observations \citep{Dahlem1993}.

\section{Observations}

The data were taken as part of programme 096.F-9312(A) using the APEX SEPIA\footnotemark  Band 5 receiver \citep{Billade2012}.
The observations were performed on 3 March 2015 at 03:40 UTC with a native channel width of 0.125 kms$^{-1}$.
The water line, of rest frequency 183.310 GHz from the para-H$_{2}$O 3$_{13}$ -- 2$_{20}$ transition, was tuned to the USB, which covered a frequency range of 
181.306 to 185.313 GHz, while the LSB covered 169.285 to 173.292 GHz. 
The observations were made in wobbler mode with an off position 140\arcsec away in azimuth and a 
frequency of 0.5 Hz. Figure~\ref{map} indicates the position and the HPBW
of the observations.

The total time on science target was 175 minutes, however the first portion of the data was discarded because of high PWV ($>$ 0.6 mm) 
leaving 88 minutes on target. In the remaining data, the water line could be seen in individual scans (5303 to 5309). 
Two later scans were discarded owing to an apparent loss of phase lock during this time. The total time remaining
on science target was 58 minutes. 

Baseline subtraction of polynomial order 2 was performed in the GILDAS CLASS package\footnotemark and the scans were averaged in time.
We used a Jy/K factor of 34 to convert between antenna temperature T$_{A}^{*}$ and flux density\footnotemark with an estimated uncertainty in the flux scale of 50\%.

\footnotetext[1]{The Digitized Sky Surveys were produced at the Space Telescope Science Institute under U.S. Government grant NAG W-2166. The images of these surveys are based on photographic data obtained using the Oschin Schmidt Telescope on Palomar Mountain and the UK Schmidt Telescope. The plates were processed into the present compressed digital form with the permission of these institutions.}
\footnotetext[2]{SEPIA is the {\it Sweden-ESO PI Instrument for APEX}, a PI instrument built in collaboration between the Group for Advanced Receiver Development, Chalmers University of Technology,  and ESO.}
\footnotetext[3]{http://www.iram.fr/IRAMFR/GILDAS}
\footnotetext[4]{https://www.eso.org/sci/activities/apexsv/sepia/sepia-band-5.html}

\begin{figure*}
  \vspace{-1.8cm}
   \centering
   \includegraphics[scale=0.45,angle=270]{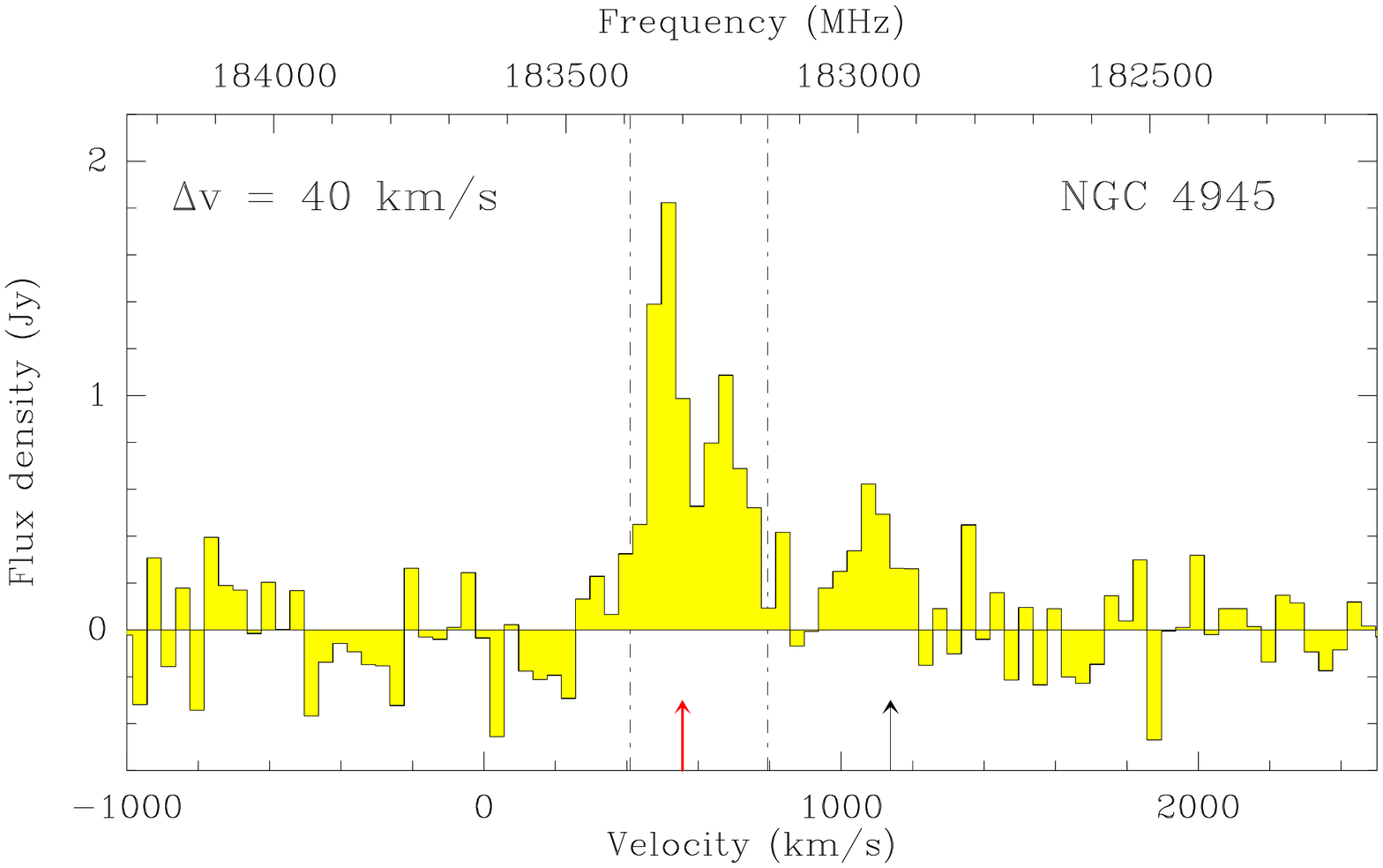}\vspace{-2.5cm}
   \includegraphics[scale=0.45,angle=270]{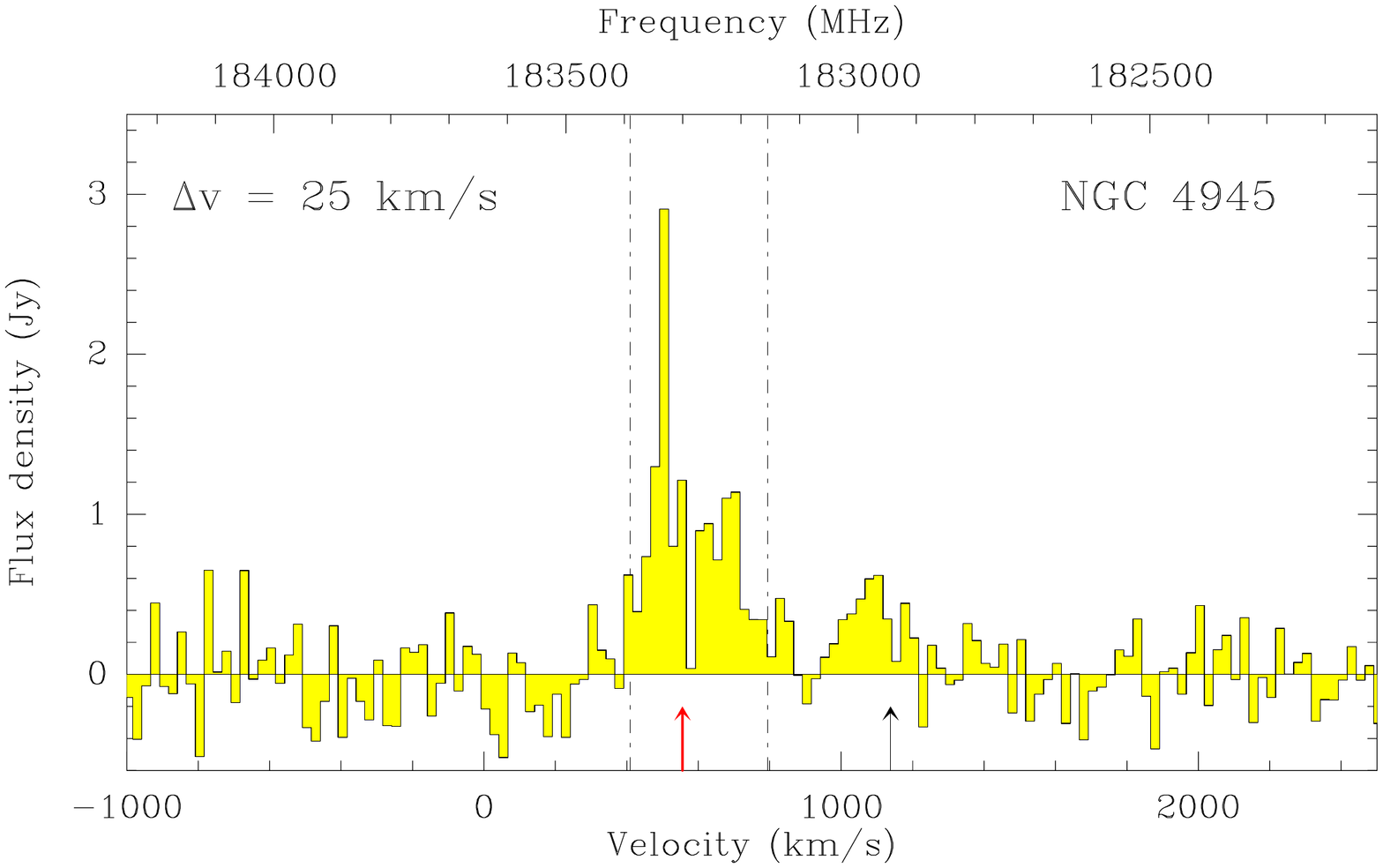}
   \caption{183 GHz H$_{2}$O emission towards NGC 4945: {\it (top)} data were binned to 40 kms$^{-1}$ resolution; {\it (bottom)} data were binned to 25 kms$^{-1}$ resolution. The red arrow indicates the approximate galactic systemic velocity
   of 556 kms$^{-1}$. The dashed lines indicate the velocity range of 22 GHz H$_{2}$O megamaser emission determined by \citet{Greenhill1997}, about 410 to 800 kms$^{-1}$. The velocities are radio heliocentric
   in \citet{Greenhill1997} and so we use v$_{LSR}$ = v$_{hel}$ - 4.6 kms$^{-1}$ to convert velocities. The black arrow denotes the velocity
   of a marginally significant feature in the 321 GHz spectrum reported by Hagiwara et al. (2016) (the strongest features at 321 GHz occurred between about 650 to 750 kms$^{-1}$). The velocity scale is radio LSR.}
              \label{spectra}%
    \end{figure*}

\section{Results}

The detection of 183 GHz water emission, obtained using APEX SEPIA Band 5, is shown in Figure~\ref{spectra} for two different smoothings. The 1$\sigma$ rms of the observations is 0.2 Jy and the peak is $\sim$1.8 Jy for the spectrum binned to a velocity resolution of 40 kms$^{-1}$. The emission is dominated by three main peaks at 505, 680, and 1084 kms$^{-1}$.
This spectrum was binned to emphasise the weaker redshifted feature at 1084 kms$^{-1}$. The 1$\sigma$ rms of the observations is 0.24 Jy and the peak is $\sim$3 Jy for the spectrum binned to a velocity resolution of 25 kms$^{-1}$. This binning is shown to indicate that the emission peak is stronger when less smoothing
is applied, which is typical for maser spectra in which individual strong spikes could have FWHM as narrow as $\sim$1 kms$^{-1}$.

\section{Discussion}

\subsection{Nature of the detected emission}

Isotropic luminosity is not likely to be a meaningful physical quantity when it comes to maser emission, 
which can be highly beamed; however traditionally this has been used for megamaser classification purposes. 
The isotropic luminosity of the 183 GHz line is given by

\begin{equation}
\frac{L}{\rm \left[ L_{\odot} \right]} = 1.04 \times 10^{-3} \frac{\nu_{rest}}{\rm \left[ GHz \right]} \frac{D^2}{\rm \left[ Mpc^2 \right]} \frac{\int S dv}{\rm \left[ Jy km s^{-1} \right] }
,\end{equation}

\noindent where $\nu_{rest}$ = 183.310 GHz.
Performing a single Gaussian fit to the blended maser emission between 400 and 800 kms$^{-1}$ gives an estimate of the integrated line
area of 11.606 K kms$^{-1}$ or 394.6 Jy kms$^{-1}$. Using a distance to NGC 4945 of 3.7 Mpc, then this gives an isotropic
luminosity of 1029 L$_{\odot}$. If emission from the redshifted feature at 1084 kms$^{-1}$ is added in, then this becomes
1303$\pm$652 L$_{\odot}$ where the uncertainty quoted is wholly based on the uncertainty in the fluxscale.

A water megamaser should have an isotropic
luminosity of a million times greater than a typical Galactic star formation water maser, and at 22 GHz this is 10$^{-4}$ L$_{\odot}$ \citep{Lo2005} (and likely less at 183 GHz), 
in which case the 183 GHz emission observed towards NGC 4945 certainly falls into the megamaser category. Indeed, famous 22 GHz megamasers towards
NGC 4258 and NGC 1068 have isotropic luminosities of 120 and 450 L$_{\odot}$ respectively \citep{Lo2005}. The isotropic luminosity 
for the 183 GHz transition found here is much higher than that found for 321 GHz emission towards NCC 4945 at $\sim$ 10 L$_{\odot}$ (Pesce2016; Hagiwara2016).
We speculate that this is due to the very different energy requirements of the two lines. Whereas the upper level of the 183 GHz transition lies at an E$_{u}$/k = 205 K
above ground state and the 321 GHz line originates from a transition at E$_{u}$/k =  1862 K. The difference between the isotropic luminosities may indicate that the physical
conditions needed to pump the 183 GHz water maser are more prevalent in the NGC 4945 central region than the conditions required to produce strong 321
GHz water maser emission. For comparison, the 22 GHz line arises from a transition of E$_{u}$/k = 644 K.

\subsection{Does the emission originate from the AGN central engine?} 

There are now three extragalactic targets reported in the literature with 
detections of 183 GHz water maser emission: NGC 3079, which is attributed to
arising from an AGN central engine as for the 22 GHz emission  \citep{Humphreys2005}; 
Arp 220, which is  attributed to a starburst origin (Cernicharo et al. 2006; Galametz et al. 2016); and this new detection
for NGC 4945. What is the likely physical origin of the newly detected maser emission in NGC 4945?

\citet[][G97]{Greenhill1997} mapped the 22 GHz H$_{2}$O emission towards NGC 4945  from 409 kms$^{-1}$ to 714 kms$^{-1}$
using a subset of the VLBA, although weak/marginal emission was detected up to about 800 kms$^{-1}$. It was found to originate 
from a linear structure extending about 40 mas ($\sim$0.7 pc at a distance of 3.7 Mpc), 
which is interpreted as a portion of the AGN circumnuclear disk.  G97 used the observations to estimate a black hole mass of
$M_{BH}$=1.4 $\times$ 10$^{6}$ M$_{\odot}$. Since 183 GHz and 22 GHz emission can be produced by similar sets of physical
conditions, and since the velocity range of the 183 GHz emission extends across the range detected at 22 GHz by G97, 
we believe that it is likely that the two maser lines originate from similar locations in NGC 4945, either entirely or predominantly
from the AGN central engine. This can be verified by future high spatial resolution observations, which
only ALMA can provide.

If the 183 GHz emission does originate from the circumnuclear disk, we can estimate the radius of emission of the
highest velocity feature at 1084 kms$^{-1}$, i.e. that closest to the black hole. For an edge-on disk and for emission arising
from the disk midline (the line perpendicular to the line of sight to the black hole) the linear
radius of the feature from the black hole is given by $r_{feature}= {G M_{BH}} / {\Delta v^2}$ = 0.022 pc 
or 1.6 $\times$ 10$^5$ Schwarzschild radii.
Hagiwara et al. (2016) detected marginally significant high-velocity emission at 321 GHz towards NGC 4945
at v$_{optical,LSR}$ = 1138.6 kms$^{-1}$. In the literature, 22 GHz maser spectra do not display velocities much above
1000 kms$^{-1}$ (e.g. G97; Pesce et al. 2016; Hagiwara et al. 2016).

\subsection{Physical conditions leading to the emission}

Multiple maser radiative transfer models provide conditions leading to
183 GHz H$_{2}$O maser emission that are appropriate to evolved stars and
star-forming regions. The 22 GHz maser emission associated with the AGN central engine disk is
believed to be pumped by X-rays from the central engine obliquely irradiating
portions of the warped disk, thereby causing heating \citep{Neufeld1994}. 
Assuming that this yields similar gas/dust conditions to galactic water masers, the presence of 183 GHz and 22 GHz emission 
likely indicates conditions found by \citet{Gray2016}. For the 22 GHz transition, strongest emission occurs between 
T$_{k}$ = 500 to 2500 K and n(H$_{2}$)=10$^{9-11}$ cm$^{-3}$.  The strongest emission occurs
between T$_{k}$ = 500 to 2000 K and n(H$_{2}$)=10$^{8-10}$ cm$^{-3}$ for the 183
GHz transition. We note the overlapping parameter
space for strong emission at both frequencies, and also that \citet{Gray2016} did not investigate densities below
10$^{7}$ cm$^{-3}$.

For Arp 220, in which 183 GHz H$_{2}$O emission is observed in the absence of 22 GHz, 
\citet{Cernicharo2006} find that significantly lower gas densities must be giving rise to the emission
(or else 22 GHz emission would also be detectable) i.e. n(H$_{2}$) $<$ 10$^{6}$ cm$^{-3}$ or T$_{k}$ $<$ 40 K.
These are conditions found in Galactic high-mass star formation,  such that the Arp 220 183 GHz emission
is proposed to be associated with the Arp 220 starburst rather than the AGN nuclei.  Lack
of variability of the Arp 220 183 GHz emission when observed with single dishes supports this hypothesis,
since the variability of the maser emission for individual star-forming cores could be washed out by beam averaging
(Galametz et al. 2016).

The 183 GHz maser emission detected towards NGC 3079 has a significantly lower isotropic luminosity
and narrower velocity range of emission than the emission detected towards NGC 4945, even though both are attributed
to arising from the AGN. There may be many reasons
for the difference. However,  in the case of NGC 3079, it appears that the strongest 22 GHz emission occurs
for a velocity range where 183 GHz emission is weak or absent. From the results of \citet{Gray2016}, this may
indicate that a significant proportion of the gas in the NGC 3079 central engine is too dense for strong 183 GHz emission to arise, while
it can still produce strong 22 GHz emission \citep[][Figure 2]{Humphreys2005} i.e. n(H$_{2}$) $>$ 10$^{10}$ cm$^{-3}$. In this
scenario, the NGC 4945 AGN would have more molecular gas in the density range n(H$_{2}$)=10$^{8-10}$ cm$^{-3}$ than NGC 3079.

\section{Conclusions}

We have made a first detection of 183 GHz H$_{2}$O maser emission towards NGC 4945 using APEX SEPIA Band 5.
The emission is strong, with a peak flux density of $\sim$3 Jy near the galactic systemic velocity.
The velocity range of the emission extends across several hundreds of kms$^{-1}$, from about 400 to 1100 kms$^{-1}$. 
The isotropic luminosity of the line classifies it as a megamaser.

Comparison of the 183 GHz spectrum with that observed at 22 GHz suggests that at least the bulk of the 183 GHz emission arises from a similar location in NGC 4945. 
From a VLBA map of the 22 GHz emission towards NGC 4945, we believe that 183 GHz emission is arising from the AGN central engine from the circumnuclear disk.
Detection of high-velocity emission at 1084 kms$^{-1}$ (528 km s$^{-1}$ from the galactic systemic velocity), if also associated with the disk, is closer to the black hole than 22 GHz emission detected to date for this target. For Keplerian rotation, the emission would be at only 0.022 pc from the black hole (1.6 $\times$ 10$^5$ Schwarzschild radii).

Further study of this megamaser should be made using ALMA Band 5. On 15 km baselines, an angular resolution of 23 milliarcseconds could be achieved. Indeed APEX now provides an ideal 183 GHz survey instrument for performing pathfinder observations
of megamaser candidates, which can then form the groundwork for spatially resolved ALMA studies.

\begin{acknowledgements}
      We thank the staff of the APEX Observatory for performing these observations. We also thank staff from Onsala Space Observatory, Chalmers University,
      and from ESO for assistance with the data. 
\end{acknowledgements}

\bibliographystyle{aa}
\bibliography{humphreys_accepted_astroph.bib}

\end{document}